# A Theory of the Big Bang in McTaggart's Time

# 1 Introduction

Exploring the issue of time as it relates to the Big Bang is difficult. On the one hand it's not clear if it is possible that 'time started' at the Big Bang. On the other hand it's not clear if it is possible that 'the universe was always here,' in which case the universe has existed infinitely far into the past. This paper gives a novel approach to time as it relates to the Big Bang. I make use of McTaggart's distinction of two series that characterize one dimension of time. There is an A-series: future-present-past, and there is a B-series: earlier-times to later-times. These two series can be varied independently, depending on the situation, (see Sect. 3). Therefore they *cannot* be regarded as the same series.

The consistency of a 'present moment' with Special Relativity is a classic problem for Presentism. An A-series has a 'present moment' in it. To accommodate a present moment this paper will propose a fragmentalist interpretation of quantum mechanics, one in which each quantum system forms a fragment in the sense that reality is divided up into these fragments. In each fragment time is described by both an A-series and a B-series. However, to accommodate Special Relativity it will be proposed that each fragment does *not* contain any information about the A-series of another fragment. That is the crucial point. So as not to make this paper unwieldy I will keep these proposals as narrowly focused as possible as they relate to a suggestion about how to understand time as it relates to the Bing Bang.

The eventual proposal will be that the Big Bang was finitely far *earlier* than the present time, yet it was infinitely far *in the past* of the present. This is just one model among a large class of models that respect the distinction between an A-series and a B-series.

2 McTaggart's Two Temporal Series Characterize One Dimension of Time

McTaggart (1908) identified two different series that may characterize one dimension of time. There is the B-series and the A-series.

> "Positions in time, as time appears to us *prima facie*, are distinguished in two ways. Each position is Earlier than some, and Later than some, of the other positions [the B-series]. And each position is either Past, Present, or Future [the A-series]. The distinctions of the former class are permanent [for time-like separated events], while those of the latter are not. If M is ever earlier than N, it is always earlier. But an event, which is now present, was future and will be past."

I will not follow McTaggart to the conclusion that time is unreal, but suppose that time is real and one dimension of time has both B-series and A-series characteristics. This is usually called an A-theory of time.

A-theorists argue the A-series is not reducible to the B-series *in any way*, and that it is a part of a comprehensive view of time. For a discussion of this see

Markosian et al. ([2017](#)). The presentist A-series of this paper consists of 1. The future, 2. An 'ontologically privileged' present moment or 'now,' 3. The past, and 4. An irreducible notion of 'becoming'. An object or event in the future *becomes* into the present and then *becomes* into the past. The B-series is an ordering that runs from earlier-times to later-times.

The idea of bringing the A- and B- series into physics has been explored in earlier papers, for example in Weingard ([1972](#), pp. 119–121), … (2019), Rovelli ([2019](#), pp. 1325–1335), and Saudek ([2019](#), pp. 1–28).

For the theory of time of this paper, instead of supposing that 'time goes from past to present to future' as McTaggart originally proposed, it would be more accurate to say that 'time goes from earlier times to later times as it becomes from the future into the present and then into a past'0.1[1](#) As later and later times become present, time goes on. This is represented in Fig. [1](#).

**Fig. 1**

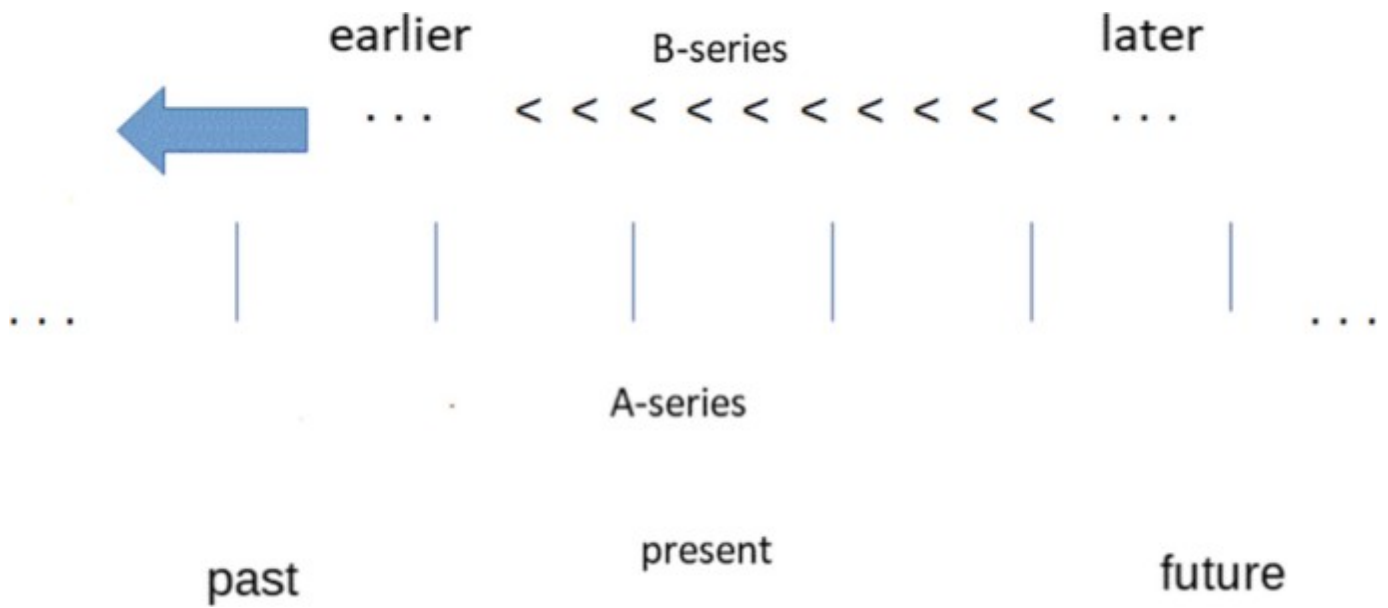

As later and later B-series times become from the future into the present and then into the past in the A-series, time goes on. This accords with experience

As later and later B-series times *become* from the future into the present and then into the past in the A-series, time goes on.[2] This accords with experience.

The next idea will be that each quantum system forms an ontological 'fragment' (Sect. [4]) that has an A-series that is 'ontologically private' to that system and delineates the fragment, but retains the ontologically public (but relativized) B-series interrelations. The point is that, while one dimension of time in one fragment is characterized by having both an A-series and a B-series, this fragment does *not* contain information about the A-series of another fragment. The A-series of each system includes an ontologically private present moment, or 'now,' that extends throughout space in the spatial coordinates of that fragment. Presumably, the *apparent* 'universal now' that extends throughout space and seems to contain all quantum systems results from some kind of averaging over the near-ubiquitous private nows.

3 Rates of Temporal Flow

I will be useful to define a 'rate of temporal flow.' It seems like 'time passes'. If it does pass, a natural question is 'how fast' does it pass? The debate often centers around the problem of what sense, if any, can be made of a 'rate' of 'one second per second.' For various points of view see, for example, Skow (2011, pp. 325–344), Prosser (2013, pp. 315–327), Maudlin (2017, pp. 75–79), Maghsoudi (2019, pp. 237–257), and Miller et al. (2020, pp. 255–280). This section of the paper will propose a novel definition of the rate(s) of temporal flow for the purpose of applying it to the question of the Big Bang. The idea will be that a rate $r$ measures 'how fast' the B-series 'goes past' the A-series present moment, for each fragment.

One may start with a parameter $t$ that models the B-series which is coordinatized by units of seconds. A B-series time $t_2$ may be 1 s later than a time $t_1$. Add a real-numbered parameter $\tau$ that models the A-series which is to be coordinatized by units of $e$s (defined below). ($e$ is not the electric charge in this paper.) $\tau$ is the future-present-past series of a given fragment (For a recent and related development see Smolin et al. 2021). Conventionally we will say that positive $\tau$ denote times in the future, a time of $\tau = 0$ denotes the present, and negative $\tau$ denote times in the past, of each fragment.[3] The idea is that *es* coordinatize *τ* the way seconds coordinatize *t*. Then an A-series time $τ_2$ can be 1 *e* further into the future (or past) of an A-series time $τ_1$.

In this theory we would say 'an event $s_1$ is 10 s later than 2 pm'. But in this model, supposing it is now 2 pm, we would *not* say '$s_1$ is 10 s in the future of 'now''. Instead, we would say '$s_1$ is 10 *e* in the future of 'now''.

The countdown to a rocket liftoff, 10… 9… 8… could be seen as counting the number of seconds later than the current time that the liftoff is, if it going to happen. It could, in addition, be seen as the number of *e*s in the future of the present that the liftoff is. In this case, when the announcer says '10' this means that the liftoff, if it is going to happen, is 10 s later than the clock-time 2 pm at the control center, and 10 *e* in the future of the present of the control center, in the conventional coordinatization. When the announcer counts down to the number '9', this means that the liftoff, if it is going to happen, is 9 *e* in the future of the present of the control center. However, the beginning of the countdown is still 10 s earlier than the liftoff—it's just that 1 s has become 1 *e* into the past.

**3.1** definitions and rates

Define an *indexical clock* to be a clock that is not accelerating, has relative velocity 0 m-per-second, and is spatially local, relative to a centered inertial reference frame, all in terms of a B-series.

Define

**3.1.1** 1 *e* is what becoming is like for 1 s of indexical clock time, without psychological distortion

The idea, which can only be touched on here, is that the A-series is (or is like) qualia. If becoming is phenomenal in the way that qualia are, then, it could be argued, it must be 'defined' or 'referred to' in this curious 'what it is like' way, as in the celebrated paper of Nagel (1974, pp. 435–450). For a more recent formulation see Farr (2020). For example a green quale is defined as 'what it is like' to experience green. The necessity of doing this has to do with qualia's ineffability. In this way *e* is well-defined in each fragment.

Note that a second is well-defined across systems such as a human, Alice, and a protozoan, even though the protozoan doesn't have the mental capacities that Alice does. It's plausible that it's the same way with an *e* of A-series time. Just as one *e* can be defined in a macroscopic fragment 'Alice,' one *e* can be defined in the fragment of, for example, a protozoa. Extending this logic we are lead to a kind of panpsychism: each quantum system, no matter how small or simple or non-local (non-local in the spatial coordinates of a different fragment), has its own A-series, and this delineates its fragment (see Sect. 4). For the rest of this paper 'Alice' and 'Bob' will refer to any quantum system (fragment) whatsoever.

Just the way one can redefine seconds to be longer or shorter than the usual seconds, one can redefine *e*s to be further or closer into the future (or past) than the usual *e*s. Of course the physically significant things should be invariant under these coordinate changes.

We can then define a rate *r* of temporal flow, for a given fragment, as.

**3.1.2** $r = -d(\text{Alice's B-series})/d(\text{Alice's A-series})$, in units of seconds/*e*

This is the change in B-series seconds per change in A-series *e*.[4][4] For example, the difference in the position of a particle at 1 s later than t = 0 may also correspond to a difference of 1 *e* closer to the present from the future (or further into the past from the present) relative to some event in the given fragment. The minus sign accounts for the fact that increasing B-series times (in seconds) become into decreasing A-series times (in *e*s), given the conventional coordinatization mentioned above. This is represented in Fig. 2.

**Fig. 2**

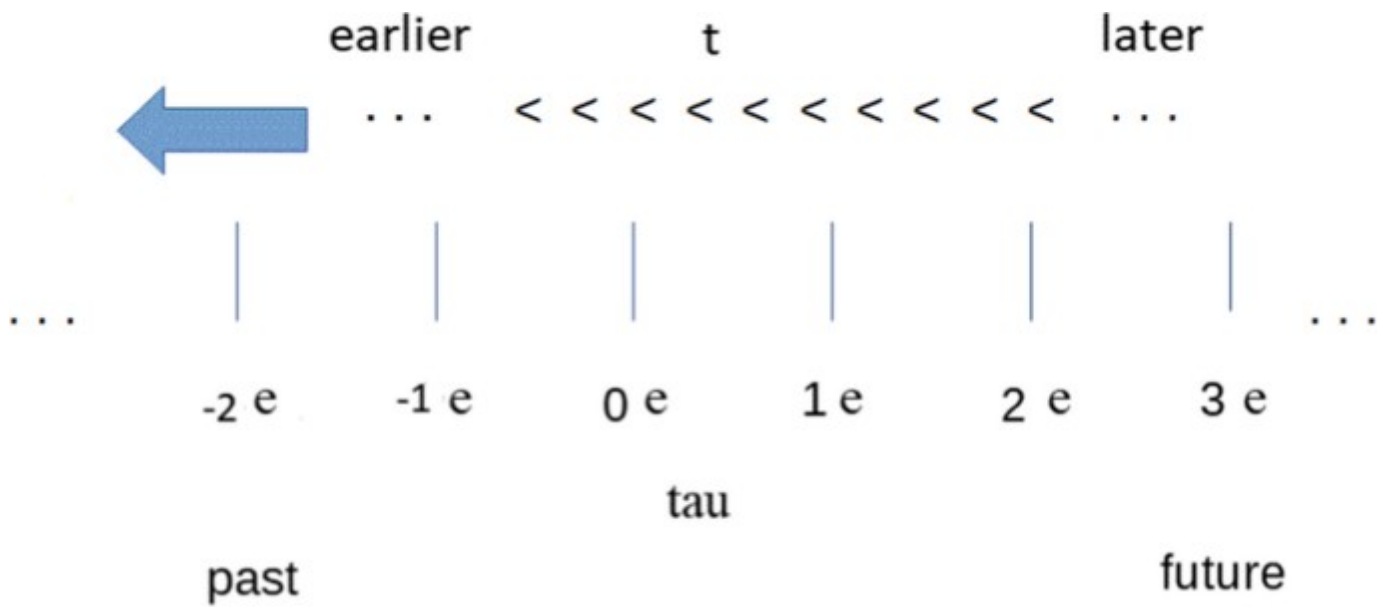

Earlier-to-later times become from the future (positive *e*) into the present (*e* around 0) and then into the past (negative *e*)

Full size image

Consider the rate $r = 2$ s/*e*. This can be interpreted as meaning there are 2 s of indexical clock time that go by per unit of becoming. That would imply that, for 1 *e*, 2 s go by. In this case earlier-to-later relations would appear to go by faster (twice as fast). This would correspond to the 'sped up movie' metaphor—one in which time goes by at twice the normal rate.

In general we have.

### 3.1.3

- $r > 1$: B-series time is experienced sped up: earlier-times to later-times are experienced as going by faster than normal.

- $r = 1$: the change in B-series information per change in A-series information is given by 1 s of indexical clock time per unit $e$ of becoming.

- $0 < r < 1$ B-series time goes by 'slowed down' relative to the A-series.

- $r = 0$: B-series time appears stopped (but the *experience* goes on as usual in the A-series).

- $r < 0$: time appears to be going backward in B-series time: later times become to earlier times (from future to present to past). This is a kind of time-reversal, and is like watching the movie go backward.

One may define (for example) $dr/de$ which would have something to do with the rate of becoming accelerating through the A-series. $e^{-2}$ would be something like 'per unit of becoming, per unit of becoming'. Etc.

Let $x$ be the position of a classical point particle defined relative to a chosen spatial origin in a given set of coordinates. One may define $dx/dt$, the 'rate' at which the position of the particle changes with respect to the B-series time $t$. This is the rate of change of the position $x$ as the B-series goes from earlier times to later times. This has units of meters/second. In the theory of this paper, one may also consider a position $y$ as a function of the A- temporal series, $y(\tau)$.6 Then $dy/d\tau$ is the rate at which the position of the particle changes as it becomes from the coordinate system's future into the system's present and then into the system's past. This has units of meters/$e$. In light of the rate $r$, $dy/d\tau$ need not equal the rate $dx/dt$. For the general case we want to define a spatial coordinate $z$ as a function of both series: $z(\tau, t)$.

For example, in high school we learn to plot the position $x$ of a classical point-particle as a function of one series of time, i.e. we plot $x(t)$. And here $t$ is a B-series. But we can also define a spatial coordinate another $y$ and plot $y(\tau)$, where $\tau$ is an A-series. In this latter case $y(5)$ means the position $y$ at 5 $e$ in the future (which might be wholly or partially in the present, depending on the duration of the present in the generalized theory). $y(0.1)$ means the position $y$ at 0.1 $e$ in the future(/present). $x(-2)$ means the position at 2 $e$ in the past(/present). Ultimately we would want to plot a spatial coordinate $z$ that is a function of the more complete notion of time, namely $z(\tau, t)$. Or, with a 'total time' $T(\tau, t)$ defined, we would plot $z(T)$.

With the definition of the rate $r$ in hand we are led to a generalization of the Lorentz transformations from the perspective of each fragment. This paper will not explore the generalizations, which necessarily involve both relativity and quantum mechanics. For a detailed discussion on the conceptual resolution to the tension between presentist fragmentalism and the relativity of simultaneity see Sect. 6.

One interesting feature of this interpretation is that the enlarged 'presentist spacetime' is, within each fragment, described by *five* parameters: an A-series parameter $\tau$, a B-series parameter $t$, and the three space parameters $x^a$. This contrasts with the *four* parameters of Minkowski space:

a B-series parameter *t*, and the three space parameters $x^a$. The development of this part of the theory must also be left to another paper. Here we just note that this definition of the 'rate of time's passing' allows a new investigation of what happens as one goes 'backward in time' toward the Big Bang.

4 Fragmentalism/Perspectivalism

The A-series of quantum systems can be accommodated within a 'presentist fragmenalist' interpretation of quantum mechanics. The idea is that each quantum system, no matter how small or simple or non-local (non-local in the spatial coordinates of another fragment) forms a fragment. Within this fragment one dimension of time is characterized by both an A-series and a B-series, and each fragment has the three spatial coordinates $x^a$, which in some sense extend throughout all of space. But the point is that the fragments are delineated by the system's A-series: one fragment does *not* contain the information of another fragment's A-series. The idea is that reality is ontologically divided up into 'fragments' this way.

There have been several interpretations of the 'fragmental' notion in physics. I'll briefly note three of them and then propose a fourth. Fine ([2005](#), pp. 261–320) introduced the notion of fragmentalism into physics. In Fine's original fragmentalism, it is supposed that different relativistic frames of reference form different fragments, so that what is true in a rest frame is not necessarily true in a frame with a (relative) velocity. 'Reality' is divided up this way. This is problematic, as Hofweber et al. ([2016](#), pp. 871–883) note, as different relativistic frames of reference can be accommodated within one ontological 'fragment' (related by the Lorentz transformations) *together*.

A second notion of fragmentalism is applied to quantum mechanics in the following sense. If Schrodinger's Cat, for example, is (schematically) in the state.

**4.1** [psi > = [alive > + [dead >

then each of the basis vectors [alive > and [dead > form two different fragments, according to Simon (2018, pp. 123–145). But this is also problematic, as Iaquinto et al. (2020, pp. 693–703) argue. It cannot be that each basis vector forms a fragment, since in the fragment of a reference (or, in this case, a 'laboratory') quantum system the state of the Cat is given by the *one* vector [psi > , which is given by the sum of the two basis vectors [alive > and [dead > taken *together*.

A third notion of fragmentalism is to posit that positions in the A-series each form different fragments, as in Torrengo et al. (2019), or, in other words, form different perspectives, as in Ludlow et al. (2016, pp. 49–74), and Slavov (2020, pp. 1398–1410). This may be tenable, and is compatible with the proposal of this paper, but I will reserve the word 'fragmental' for a different sense.

The notion of fragmentalism used in this paper is that each quantum system (no matter how small or simple or non-local) forms a fragment. In one fragment there is no fact of the matter about the value of the relevant parameter in another fragment. In this paper I will take the relevant parameter to be the A-series. Thus there is one separate A-series for each quantum system.

Each fragment has an A-series, a B-series, and a (relativistic) B-series of another fragment. But, again, a first fragment does *not* contain the information of the A-series of a second fragment. And vice versa. That is the crucial point for the theory. Reality is divided up into fragments in this way. The A-series parameter in more than one fragment is simply undefined.

We assume one fragment is just as good as any other fragment from which the laws of physics must hold. In the presentist fragmentalist interpretation here, two fragments become to have the same A-series when and only when the two quantum systems 'observe' or 'measure' or 'collapse the state-function' of each other. This is described as usual by an operator on a Hilbert space as defined from each fragment.

In this notion of fragmentalism we suppose that any one quantum system can serve as a reference system just as well as any other quantum system. Then system $q_1$ serves as a reference system whose parameters describe another system $q_2$ if and only if there is a reciprocal description of $q_1$ in the parameters of the system $q_2$. See …(2005), Giacomini et al. (2019).

There is a great deal more to say about presentist fragmentalism but it would take us beyond the bounds of this paper. Here I stick to the ambition of narrowly applying these ideas to a model of the Big Bang.

## 5 The Big Bang

It may be possible to apply the ideas of Sects. 2–4 to model the Big Bang. First I will consider two extremal cases that involve both the A-series and the B-series.

**5.1** the Big Bang was infinitely earlier than now (in the B-series) and finitely far in the past (in the A-series)

and.

**5.2** the Big Bang was finitely earlier than now (in the B-series) and infinitely far in the past (in the A-series)

I would argue that case (5.2) is a better model than case (5.1). The argument is that we find the Big Bang to be 13.8 billion years earlier than now. But this leaves open the question of why the Big Bang did not happen a billion years before now, such that we are also a billion years before now.5 It would seem that a notion of time that is based on just the B-series cannot answer this question.

In case (5.2) the number of seconds that the Big Bang is earlier than now could remain given by 13.8 billion years. But as we go to earlier and earlier times toward the Big Bang, it could be that we have to go ever further into the past. In this case, to successively go 1 s earlier in the B-series requires going progressively to a larger and larger extent into the past (in units of *e*) as we approach the Big Bang. That is the first basic proposal about the Big Bang of this paper.

Then, in light of the definition of rate(s) in Sect. 3, case (5.2) gives.

### 5.3 $r \to 0$ s/e

As noted earlier, quantum systems form distinct fragments. In the presentist fragmentalist interpretation here, a mutual measurement/observation/collapse is given when and only when there is a (mutual) collapse of the wavefunction as defined in each respective fragment. At mutual collapse the two A-series of the respective systems become one A-series. I will not explore this part of the model here except to note that.

**(5.4)** such projections update the value of the rate $r$ for each fragment.

Together, (5.3) and (5.4) predict that as one goes back along $t$ in the B-series to ever earlier times toward the Big Bang there will be, for each second, a larger and larger number—tending toward infinity—of *e*s going into the past along $r$. This is the same as increasing the number of quantum interactions per second as one goes backward in time. That is the second basic proposal about the Big Bang of this paper.

If we take the speed of light $c = 1$ m/second as a conversion factor, then the closer we get to the Big Bang the larger the number of quantum interactions per unit 4-volume. This looks like it approaches an infinite number of quantum interactions per 4-volume of Minkowski space as we go toward the singularity. That is a basic consequence of the model of this paper.

6 Einstein's Train in the Presentist Fragmentalist Interpretation of Quantum Mechanics and its Consistency with the Relativity of Simultaneity

It is worth showing how the particular version of presentism in this paper is consistent with special relativity and, in particular, the relativity of simultaneity. We consider 'Einstein's train,' Einstein ([1905](), pp. 891–921).

Suppose Alice is standing in the middle of the platform of a train station and lightning strikes each end of the platform simultaneously in her reference frame. She knows the strikes were simultaneous in her reference frame because after each strike a classical message (perhaps on pieces of paper) is sent to her about when they happened.

Suppose Bob is standing in the middle of a train car that is the length of the platform in the rest frame of the car, and that the car is moving past the station parallel to the platform.

Of course, the point is that the two lightning strikes will *not* be simultaneous in Bob's frame of reference.

It is often concluded that there is no ontologically privileged present moment, or no unique 'now,' since if there were such a moment it would be simultaneous with itself(!), and therefore could not encompass both Alice's frame of reference and Bob's frame of reference.

We'll look at the train thought experiment in more detail in the presentist fragmentalist interpretation. The basic idea will be that Alice's system forms an ontological fragment that does indeed have a unique present moment that extends throughout all of space. She has the information of her A-series and her B-series within her fragment. She will also have the information of Bob's B-series, as modified by the Lorentz transformations in the coordinates of Alice's B-series. But her fragment does *not* contain the information of Bob's unique present moment (in particular, his A-series). And vice versa. There is just no fact of the matter about both A-series or both present moments taken together before mutual state-vector collapse.

Einstein's train reconsidered.

Alice stands at the train station in the middle of the platform, ready for the experiment to commence. She is 'always' in her ontologically privileged present moment. Arguably, this is empirically given data for her. She can talk about and think about her future and her past all she wants, but, arguably, it is only in Alice's and Bob's mutual present that she can *demonstrate* an experimental outcome. This is an important point, but the discussion must also be left for another paper.7 Suppose the train car containing Bob (who is of course at rest relative to the train car) moves past the platform.

Many events that are simultaneous in Alice's frame of reference will not be simultaneous in Bob's frame of reference, and vice versa. The B-series space and time of Bob's frame of reference gets modified by the Lorentz transformations relative to Alice's B-series space and time. Taking into account both Alice's A- and B- series there is a generalization of the Lorenz transformations for Bob's B-series.

So far, there is no contradiction. We have only talked about Alice's present moment (given by her A-series), her clock-times (given by her B-series), and the clock times of Bob's relative frame of reference (given by his B-series from the perspective of Alice's fragment).

The idea is then that Alice's A-series delineates her ontological fragment. A contradiction would arise only if we added the postulate that Bob has a unique present moment *in Alice's fragment*. But this postulate is avoided if we suppose that Bob's A-series delineates his own fragment—one that is distinct from Alice's fragment—and that its value(s) are not given in Alice's fragment.

More explicitly, Alice's fragment contains the information of Alice's A-series, Alice's B-series, and Bob's (transformed) B-series. But, to reiterate, Alice's fragment does *not* contain the information of Bob's A-series. Similarly, Bob's fragment contains the information of Bob's A-series, his B-series, and Alice's (inversely transformed) B-series. But his fragment does *not* contain the information of Alice's A-series. As they pertain to two different fragments, there is explicitly no 'simultaneous' fact of the matter about both Alice's A-series (and therefore her present moment and her 'becoming') and Bob's A-series (and therefore his present moment and his 'becoming').

In Alice's fragment, Bob's B-series is modified by Lorentz transformations relative to her B-series (as given in her fragment). And in Bob's fragment, Alice's B-series is modified by the inverse Lorentz transformations relative to his B-series (as given in his fragment).

In this way, there is a present moment that extends throughout space for each fragment, yet there is no mutual present moment for the two fragments, so no contradiction ever arises.

This concludes the Presentist Fragmentalist account of the Einstein's train thought experiment.

Epilogue to Einstein's Train.

Clearly there is a great deal more to say. But the aim of this section was only to give a narrow conceptual account of the train/station thought experiment because that is one of the stem arguments showing that simultaneity is relative, which is (erroneously) thought to imply that presentism is untenable.

At the risk of gross oversimplification, and with reference to the presentist fragmentalist theory developed in a cognate paper (… 2021) the slogan here might be that the times of quantum mechanics are (fragmentalist) A-series, and the times of relativity are B-series, with the understanding that these two temporal series are not entirely independent. The B-series of the opposing fragment gets modified by the Lorentz transformations as usual. There is a generalization of these transformations when both the A-series and the B-series from the perspective of the reference fragment are taken into account.

## 7 Conclusions

McTaggart's distinction between the two series that characterize one dimension of time allow for a distinction between the two questions: what happens as we go to *earlier times* toward the Big Bang (in the B-series)? And: what happens as we go *further into the past* toward the Big Bang (in the A-series)? There is a simple model presented in which the Big Bang was 13.8 billion years *earlier* than the present, but in which we would nevertheless have to go infinitely far *into the past* to get to the Big Bang. Of course, this is just one model out of a large class of models that respect the distinction between the A-series and the B-series.

To do justice to the theory within which this distinction can be made would take us wildly outside the bounds of this single paper. But notes on how such a theory could work were touched upon: we defined a 'rate of temporal flow' (Sect. 3); a presentist-fragmentalist interpretation of quantum mechanics (Sect. 4); two possible models of the Big Bang (Sect. 5); and explicitly show how these ideas are compatible with special relativity (Sect. 6).

The conceptual playground here allows us to state and address novel ideas about fundamental questions about the Big Bang, models of time, and interpretations of quantum mechanics.

Notes

1. McTaggart sometimes uses the A-series as past-present-future. In this paper we'll use the A-series in the opposite direction: as future-present-past. This is because of the oft-noted property that an event is first in my future, second in my present, and third in my past.

2. This diagram raises the near-ubiquitous problem of super-time. Time is supposed to be given by the A-series and the B-series. Yet the arrow in the top left of the figure is required to indicate that the B-series 'moves' past the A-series. But this 'movement' is within neither the given A-series nor the B-series. This movement would, then, have to be defined within a *third* temporal series. But if we plotted the third series in a diagram we would have to have 'movement' in *it* in some way, and that would require a *fourth* temporal series, and so on. This leads to the well-known infinite regress. Unfortunately we cannot develop the full theory of time (and indeed quantum mechanics) in this one paper. But the idea will be that such 'movement' is given by an operator that irreducibly *operates* (see 5.4). The operation is, in some sense, irreducibly a *verb*.

3. For the general case we would not necessarily assume that the present is restricted to a point at $\tau = 0$, but instead is given by some function on $\tau$.

4. 4Note that in this presentist theory a rate $r$ neither assumes not implies that the future is predetermined. The argument is that there may be many future states that are consistent with a fragment's present state. The exploration of this presentist notion is given in a related paper.

5 Note that the two cases in this argument are empirically indistinguishable only on some models of time.

6 It is intended that $x$, $y$, and $z$ in this paragraph each refer to a general position defined in a space. In particular do not refer to different axes.

7Because of this empirically given fact I would go so far as to claim that every scientist should be a presentist.

[Download references](#)

Funding

The authors have no relevant financial or non-financial interests to disclose.

Author information

**Authors and Affiliations**

1. **Santa Cruz, CA, USA**

    Paul Merriam

**Corresponding author**

Correspondence to Paul Merriam.

Ethics declarations

**Conflicts of interest**

None.

Additional information

**Publisher's Note**

Springer Nature remains neutral with regard to jurisdictional claims in published maps and institutional affiliations.

## Rights and permissions